\documentstyle[sprocl]{article}
\input{psfig}
\def\be{\begin{equation}}
\def\ee{\end{equation}}
\def\bea{\begin{eqnarray}}
\def\eea{\end{eqnarray}}

\begin{document}

\title{THE X--RAY BACKGROUND AND THE ROSAT DEEP SURVEYS}
\author{ G. HASINGER }
\address{Astrophysikalisches Institut Potsdam, An der Sternwarte 16,\\
14482 Potsdam, Germany\\E-mail : ghasinger@aip.de}
\author{ G. ZAMORANI }
\address{Osservatorio Astronomico, Via Zamboni 33,\\
Bologna, 40126, Italy\\
E-mail: zamorani@astbo3.bo.astro.it}

\maketitle\abstracts{
In this article we review the measurements and understanding of the X-ray 
background (XRB), discovered by Giacconi and collaborators 35 years ago.
We start from the early history and the debate whether the XRB is 
due to a single, homogeneous physical process or to the summed 
emission of discrete sources, which was finally settled by COBE and
ROSAT. We then describe in detail the progress from ROSAT deep surveys
and optical identifications of the faint X-ray source population. In 
particular we discuss the role of active galactic nuclei (AGNs) 
as dominant contributors for the XRB, and argue that so far there is 
no need to postulate a hypothesized new population of X-ray sources. 
The recent advances in the understanding 
of X-ray spectra of AGN is reviewed and a population synthesis model,
based on the unified AGN schemes, is presented. This model is so far
the most promising to explain all observational constraints. Future
sensitive X-ray surveys in the harder X-ray band will be able to 
unambiguously test this picture. 
}

\section{Introduction}
It is a heavy responsibility for us, as it would be for everybody
else, to write a paper on the X--ray Deep Surveys and the X--ray 
background (XRB) for this book, in honour of Riccardo Giacconi. 
Everybody recognizes the importance of Riccardo's contribution in both
these fields. The discovery of the XRB using proportional counters
in a rocket flight, the UHURU satellite, the first X--ray telescope with
focusing optics ``Einstein'' are milestones in X--ray astronomy
and Riccardo played a leading role in these experiments.
More recently he gave a fundamental contribution to the planning,
execution and analysis of the ROSAT Deep Surveys. As a consequence
of this, a significant fraction of
the results which we will describe in this paper either are
his own results or are based on experiments which he conceived and
led to success.

In Section 2 we give a brief historical overview of the XRB problem,
from its discovery up to the results obtained in the eighties 
with the HEAO--1 and Einstein missions. During these years the origin of the
XRB has been discussed mainly in terms of two alternative interpretations: 
the diffuse hypothesis (e.g. hot intergalactic gas) and the discrete
source hypothesis. The existence of these radically alternative hypotheses
has not been
``neutral'' with respect to devising experiments which wanted
to study the XRB. In fact, if the XRB were mainly due to discrete sources,
experiments aimed at studying the single sources responsible for it
would obviously need high angular resolution 
in order to detect and resolve the large number of expected
faint sources. Vice versa, if the XRB were mainly diffuse, source confusion
would not be a problem and one could safely abandon the high angular
resolution option. In this case the crucial experiment would
be a measurement as accurate as possible of the spectrum in order to reveal
the dominant physical production processes. The two working hypotheses led
various groups of scientists to design very different sets of
experiments~\cite{gb}.
In Section 3 we present and discuss some recent results from
deep surveys with ROSAT. These surveys have already resolved into
discrete sources $\sim$ 70\% of the measured XRB in the 1--2 keV band.
The available optical identifications, still in progress, suggest that
AGNs are the dominant population at these faint X--ray fluxes. 
Finally, in Section 4 we discuss some models which, taking into account
the existing observational data on AGNs, are able to produce reasonably
good fits to the XRB spectrum up to $\sim$ 100 keV.
The discussion is summarized in section 5.

\section{The Early History of the X--ray Background}
The existence of a diffuse XRB was discovered more than 35
years ago~\cite{rg62}. The aim of the experiment
was to measure the X--ray emission from the Moon, but the data showed
something unexpected: a strong X--ray source about 30 degrees away from the
Moon (Sco X--1) and a diffuse emission approximately constant from all the
directions observed during the flight.
After this discovery, the first real improvement in our
knowledge of the XRB has been made with the first all--sky
surveys (UHURU and ARIEL V) at the beginning of the seventies. The high
degree of isotropy revealed by these surveys led immediately to
realize that the origin of the XRB has to be mainly extragalactic. Moreover,
under the discrete source hypothesis, the number of sources contributing
to the XRB has to be very large~\cite{sc} ($N > 10^6 \, sr^{-1}$). 

In the same years a number of experiments were set up to measure
the spectrum of the XRB over a large range of energy. 
It was found that over the energy range 3--1000 keV the
XRB spectrum is reasonably well fit with
two power laws with energy indices $\alpha_1 \sim 0.4$ for $E \leq 25$ keV and
$\alpha_2 \sim 1.4$ for $E > 25$ keV (see Figure 1 in~\cite{ta}).

At the beginning of the eighties two different sets of measurements
led additional fire to the debate between supporters of the discrete source and
diffuse hypotheses. On the one hand, the excellent HEAO--1 data showed that
in the energy range 3--50 keV the shape of the XRB is very well fit
by an isothermal bremsstrahlung model corresponding to an optically
thin, hot plasma with kT of the order of 40 keV~\cite{mar}. 
Moreover, it was shown that essentially all Seyfert 1 galaxies
with reliable 2--20 keV spectra ($\sim$ 30 objects, mostly from HEAO--1 data)
were well fit by a single power law with an average
spectral index of the order of 0.65, significantly different from
the slope of the XRB in the same energy range~\cite{mu84}. 
Since Active Galactic Nuclei
(AGNs) were already considered the most likely candidates for the
production of the XRB under the discrete source hypothesis~\cite{sw73}, 
these two observational
facts were taken as clear ``evidence'' in favour of the diffuse thermal
hypothesis.
On the other hand, the Einstein observations were
suggesting a different scenario: 

a) Pointed observations of previously
known objects very soon showed that AGNs, as a class,
are luminous X--ray emitters~\cite{ht}.

b) When a larger AGN sample became available,
it was confirmed that AGNs could contribute most of the diffuse soft 
XRB~\cite{gz81}. Actually, in order to avoid a contribution from
AGNs larger than the observed background, it was concluded~\cite{gz81} 
that the optical counts of AGNs had to flatten at magnitudes slightly fainter
than the limit of the optically selected samples existing at that time. 
Such a flattening was later seen in deeper optical surveys.

c) Deep Einstein surveys showed that about 20\% of the soft XRB (1--3 keV) 
is resolved into
discrete sources~\cite{rg79,gr,pr,ha} at fluxes of the order of a 
few $\times \, 10^{-14} erg\,cm^{-2} s^{-1}$.
A large fraction of these faint X--ray sources were identified with 
AGNs.

Because of the difference between the spectra of the XRB and those of
the few bright AGNs with good spectral data, the supporters of the
diffuse, hot plasma hypothesis had to play
down as much as possible the contribution of AGNs to the XRB 
to a limit which was close to be in conflict with an even mild extrapolation
of the observed log(N)--log(S).
Actually, a number of papers 
were published in which it was ``demonstrated'' that even in the soft X--ray
band AGNs could not contribute much more than what had already been
detected at the Einstein limit. 

At that time we personally think that there was already evidence (for
those who wanted to see it...) that the diffuse thermal emission
as main contributor to the background was not tenable
(see, for example,~\cite{se}).
Very simple arguments in this direction were given in~\cite{gz}. 
On the basis of reasonable
extrapolations of the X--ray properties and the optical
counts of known extragalactic X--ray sources (mainly AGNs and galaxies),
Giacconi and Zamorani concluded that it is unlikely that their contribution to the
soft X--ray background is smaller than 50\%. Given this 
constraint, they then discussed two possibilities: 

i) either faint AGNs have the so--called (at that time) ``canonical'' spectrum
observed for brighter AGNs. In this case the residual XRB
(i.e. the spectrum resulting after subtraction of the contribution
from known sources)
would not be fitted anymore by optically thin bremsstrahlung;

ii) or spectral evolution for
AGNs is allowed. In this case, in order not to destroy the excellent
thermal fit in the 3--50 keV data, diffuse emission could still
be accommodated only if discrete sources have essentially the same
spectrum as the XRB. 
On this basis, they concluded that ``since in this scenario we would already
require that the average spectrum of faint sources yielding 50\%
of the soft XRB is essentially the same as the observed
XRB, there is nothing that prevents us from concluding that the
entire background may well be due to the same class of discrete
sources, at even fainter fluxes''. 

In other words, reversing the usual line of thought, 
the excellent thermal fit of the 3--50 keV 
XRB spectrum was shown by these arguments to be a point in favour
of the discrete source hypothesis, rather than of the hot gas hypothesis!
These conclusions, however, were not well received in a large
fraction of the X--ray community; probably, they had the
defect of being too simple and direct...

Thus, the debate between the supporters of the two hypotheses continued,
until the final resolution of the
controversy came from the incredibly neat results obtained with the FIRAS
instrument on board COBE: the absence
of any detectable deviation from a pure black body of the
cosmic microwave background set an upper limit to
the comptonization parameter $y < 10^{-3}$,
more than ten times smaller than the value required by the hot 
intergalactic gas model~\cite{mat90}. The most
recent upper limit for the comptonization parameter~\cite{mat94}
is now $y < 2.5 \times 10^{-5}$.
Discussing these data, in~\cite{wr} it is concluded that a uniform,
hot intergalactic gas produces at most $10^{-4}$ of the observed
XRB!

\section{The ROSAT Deep Surveys}

\subsection{Instrumental considerations}

\begin{figure*}[tp]
\unitlength1cm
\begin{minipage}[t]{5.5cm}
\begin{picture}(5.5,5.5)
\psfig{figure=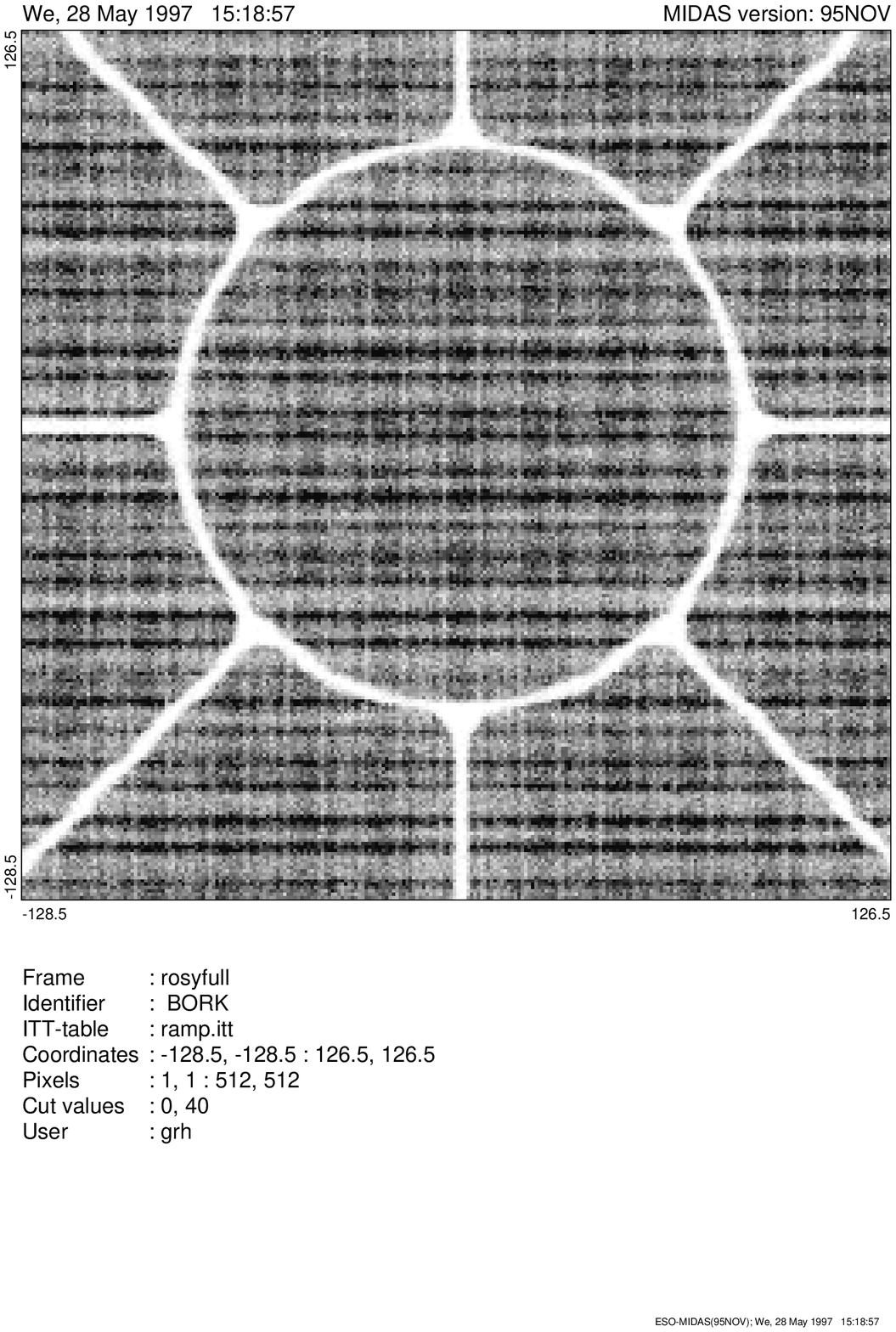,width=5.5cm,clip=}
\end{picture}\par
\end{minipage}
\begin{minipage}[t]{5.5cm}
\begin{picture}(5.5,5.5)
\psfig{figure=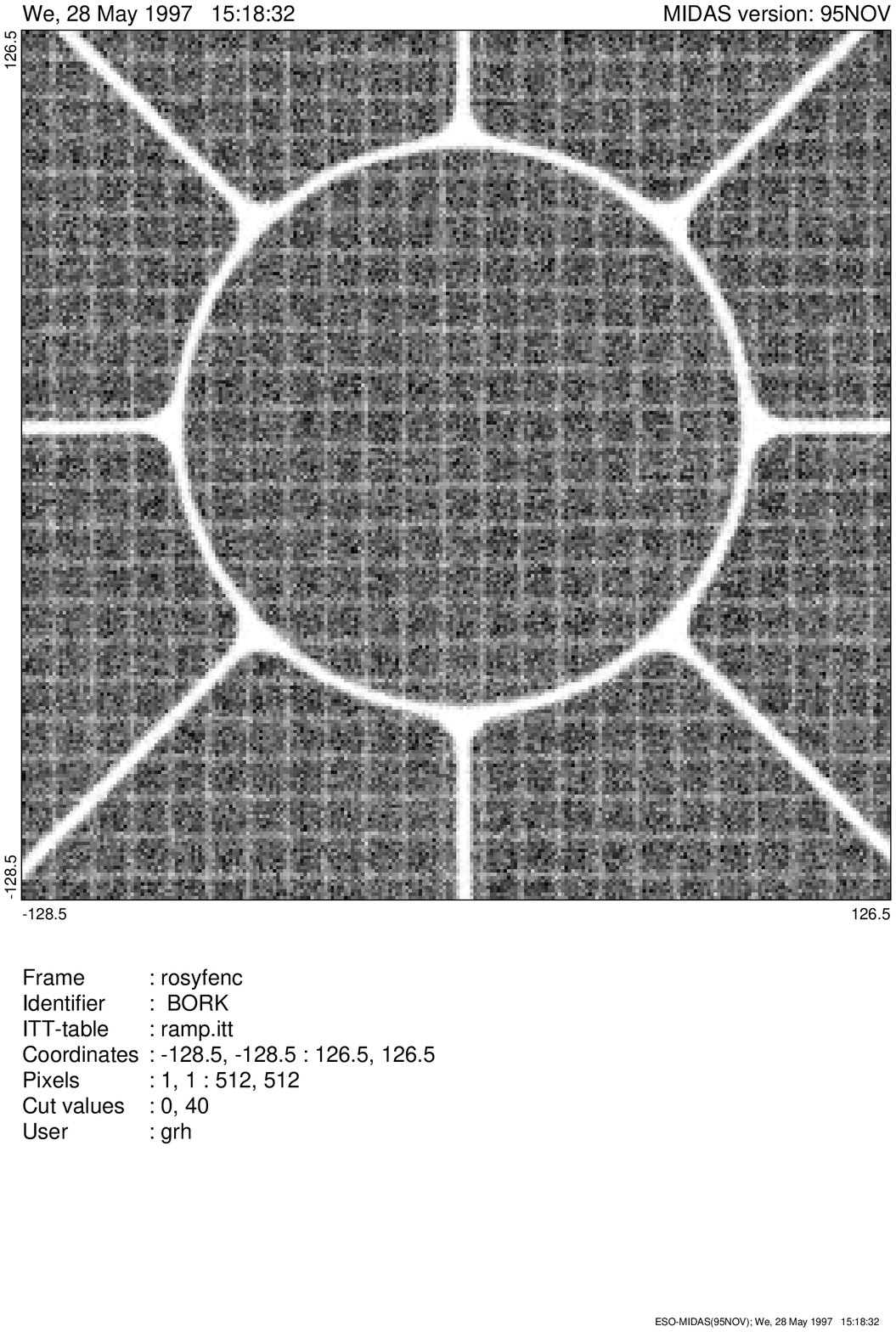,width=5.5cm,clip=}
\end{picture}\par
\vskip -0.2 truecm
\end{minipage}
\caption{\it The ROSAT PSPC flat-field in detector coordinates.
Left: uncorrected; right: corrected for geometric distortions
(the web-like shadow is due to the mechanical support structure
of the PSPC window).}
\label{PSPCFF}
\end{figure*}

The detailed preparation for ROSAT Deep Survey observations
started in 1984, well in advance of the actual ROSAT mission. At
this time the engineering model of the ROSAT Proportional Counter
PSPC~\cite{pfe86} had already been calibrated in the
laboratory and the task was at hand to understand and try to
correct for the various electronic and geometric distortions and
gain non--linearities, which in general plague imaging proportional
counters. Non--linearities in the performance of the Einstein IPC
have e.g. set the final sensitivity threshold for Einstein Deep
Surveys~\cite{sol91,ham92}. The ROSAT PSPC, even at the
raw coordinate level, shows a higher degree of uniformity than the
IPC. It was therefore possible to largely correct the significant
distortions present in the PSPC data~\cite{bri88}. Figure
\ref{PSPCFF} shows the comparison of a PSPC ground calibration flat field
before and after correction of the distortion effects.

These images immediately indicate another geometrical problem.
While it is obviously possible to remove the distortions created
by the detector itself, the shadows cast by the complicated 
wire-mesh in front of the PSPC cannot be corrected for. The mere fact
that the 100$\mu$ wire shadow can be detected, demonstrates the
relatively good angular resolution of the PSPC. In the early days
we were still hoping that the satellite pointing instability
would wash out any residual flat-field inhomogeneities.

Roughly at the same time we started prototyping the ROSAT
scientific analysis software. Early ideas about local and map-detect
algorithms, background estimation etc. were taken over from
the Einstein analysis system. However, substantial improvements were
incorporated, the most important of which was probably the
rigorous application of Poisson statistics and maximum likelihood
estimators in all statistical computations.

In order to test and understand the analysis algorithms we 
developed a science simulator system, which turned out to be one
of our most powerful tools in the preparation of the science
mission. During the course of time the simulation models became
more and more realistic, including extended and point sources,
time variability, different spectral models and realistic number
counts for the X--ray sources as well as cosmic and solar
scattering backgrounds. Orbital variation of the exposure time and
background components due to the radiation belts were included as
well. Every individual photon was traced through a realistic
model of the X--ray mirror system and the detector~\cite{har86}.
Comparing input and output information, the detection algorithms
could be tested, calibrated and, if necessary, improved.

\begin{figure}[tp]
\unitlength1cm
\begin{center}
\begin{minipage}[t]{5.5cm}
\begin{picture}(5.5,5.5)
\psfig{figure=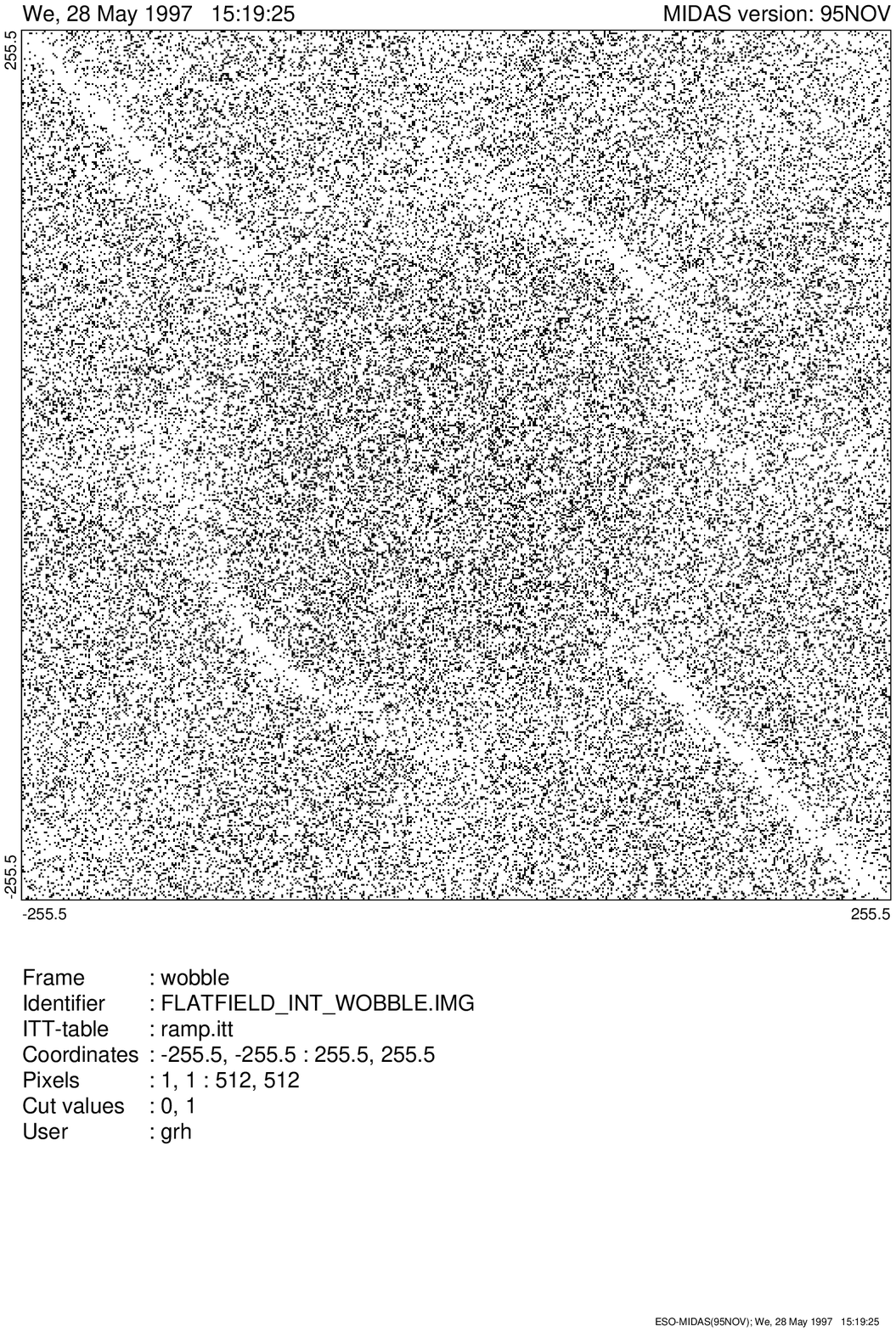,width=5.5cm,clip=}
\end{picture}\par
\end{minipage}
\end{center}
\caption{\it The ROSAT PSPC flat-field in sky coordinates
after the application of the wobble mode.}
\label{WOBBLE}
\end{figure}

In 1985 the first data about the performance of the
ROSAT attitude control system became available from dynamical
hardware-in-the-loop tests. It turned out that the attitude
control was much more stable than specified and anti\-cipated.
Simulations including realistic attitude data indicated that sources in
the center of the PSPC field could easily get lost behind
shadows of the PSPC support grid. Less than a year from the
originally planned launch (1987) we were able to convince the
funding agency and industry to introduce a ``wobble mode'' into the
attitude control system, which was to become the standard ROSAT
pointing mode. Instead of wandering around at random, the
spacecraft pointing direction is guided smoothly into a linear
periodic zig-zag motion with a period of 402 seconds and an
amplitude of $\pm 3$ arcmin ($\pm 1$ arc\-min for the HRI) diagonal with
respect to the detector coordinates. The wobble-mode, while introducing
artificial periodic power into the measured light curves of X--ray
sources, turned out to be very valuable for the study of the X--ray
background or other diffuse sources. Figure \ref{WOBBLE} shows the PSPC
flat-field expected after the wobble motion has been applied to
it. The rms variation of the wobbled PSPC flat field is about 
3 percent~\cite{has93}. 

\subsection{Science preparation}

In 1985 we assembled an international team of astronomers
interested in deep X--ray surveys. Riccardo Giacconi, together with
Richard Burg, at that time both at STScI, brought the Einstein
experience into the team and originally suggested the deep survey
in the Lockman Hole as our prime study area. This hole had just been
discovered as the direction with an absolute
minimum of interstellar hydrogen column density~\cite{loc86}. 
Gianni Zamorani brought the experience and the data of the deep
optical survey in the Marano field~\cite{mar88}, which we defined as a second
study area in the southern hemisphere. Maarten Schmidt, who had
previously modeled~\cite{sgr83} the contribution of various classes of
sources to the X--ray background, was planning the optical identification
work in the North, first using the Palomar 200" and later the Keck
telescopes. Joachim Tr\"umper, Gisela Hartner and G\"unther Hasinger
were responsible for securing the ROSAT data, i.e. calibration,
simulation and finally observation and analysis. Since then the
group met regularly at 1/2-1 year intervals. We  recall these
meetings as extremely productive but also quite exhaustive and
often with violent disputes. In particular we all appreciated
Riccardo's rigorous and constructive criticism.

The Lockman Hole is actually several degrees across. 
At first we chose a $\sim 0.3~deg^2$ sky region close to the 
absolute minimum of $N_H$, but away from bright stars.
Well in advance of the actual X--ray observations we performed 
optical UBVRI and radio observations in this field. In 1988 a mosaique 
in U, B and R
was taken with the University of Hawaii 2.2m telescope and in 1989 drift 
scans in V and I were performed with the 4-shooter camera at the 
Palomar 200"-telescope. A 16 hr mosaique of observations~\cite{der97} at 
20 cm in the VLA D-configuration was taken in 1990. 
In the meantime other groups have picked up our fields as deep 
study areas. Both the Lockman Hole and the Marano field are covered 
by deep ISOCAM FIR imaging studies (Cesarsky et al.), the Lockman 
Hole is e.g. targeted in the Heidelberg CADIS NIR survey, in 
Luppino's wide field weak lensing survey and in ASCA deep 
survey observations. 

\subsection{ROSAT observations}

Deep X--ray survey observations in the Lockman Hole commenced in the 
ROSAT AO-1 (spring of 1991) with a 100 ksec PSPC pointing exposure.
The PSPC exposure was 
aimed from the beginning to reach the ultimate instrumental limits,
while an HRI raster scan was planned to improve the PSPC positions. 
Since the spatial homogeneity of ROSAT observations is very good (see above), 
their ultimate sensitivity limit is set by confusion. In order to fight 
confusion, we had to obtain a very good understanding of the PSPC 
point-spread function~\cite{has94}, develop a completely
new algorithm for X--ray crowded field analysis, and 
run large numbers of simulations in order to understand the systematic 
subtleties and the limitations of the observations. 
A substantial fraction
of this work was already available before the actual observations. The 
knowledge about instrumental limitations has led us to the final choice
of the exposure time and to a conservative selection of flux limits and
off-axis ranges for the complete samples to be analyzed with PSPC data. 

Because of the expected confusion in the PSPC, it was also clear from the 
beginning that HRI data would be necessary to augment the PSPC identification
process. Because of the HRI smaller field-of-view,  
lower quantum efficiency and higher intrinsic background, we figured that it 
was necessary to invest about a factor four more HRI time than PSPC time to 
cover the same field with the same sensitivity, based on pre-launch knowledge.
We therefore started a raster scan in AO-1, with $\sim 100$ pointings of 2 ksec
each 
across the survey region. The remaining 200 ksec of HRI raster observations were
approved in AO-2. When, however, the first HRI in-flight performance figures
became available~\cite{dav96}, we realized that the anticipated sensitivity 
would not be reached with the HRI raster scan. Due to the increased quantum 
efficiency,
compared to the Einstein HRI, the ROSAT HRI is also more susceptible to 
background induced by particles in orbit. An increased halo of the HRI 
point-spread function as well as irreproducible attitude errors of about 5" 
are responsible for a further loss of sensitivity. Knowing this, we were 
able to trade the 200 ksec HRI time remaining for AO-2 into an extra PSPC
observation of 100 ksec, which was performed in spring of 1992.   

Two years later, after the PSPC observations of the Marano field had been
successfully completed and the PSPC had run out of gas, we started to apply
for an HRI ultradeep survey aimed for a total observing time of 1 Msec in a
single pointing direction. This survey
was planned to push the unconfused sensitivity limit deeper than the PSPC 
exposure in a substantial fraction of the PSPC field.
In order to allow an X--ray ``shift and add'' procedure, correcting for the
erratic ROSAT pointing errors, we selected a pointing direction for the 
ultradeep HRI exposure which is inside the PSPC field of view, but shifted about
8 arcmin to the North--East of the PSPC center, this way covering
a region containing about 10 relatively bright X--ray sources known from
the PSPC and the HRI raster scan. The 1 Msec HRI survey was approved for 
ROSAT AO-6 and AO-7 and is still going on. So far an exposure time of 
880 ksec has been accumulated.
 
\begin{figure*}[tp]
\unitlength1cm
\begin{minipage}[t]{5.5cm}
\begin{picture}(5.5,5.5)
\psfig{figure=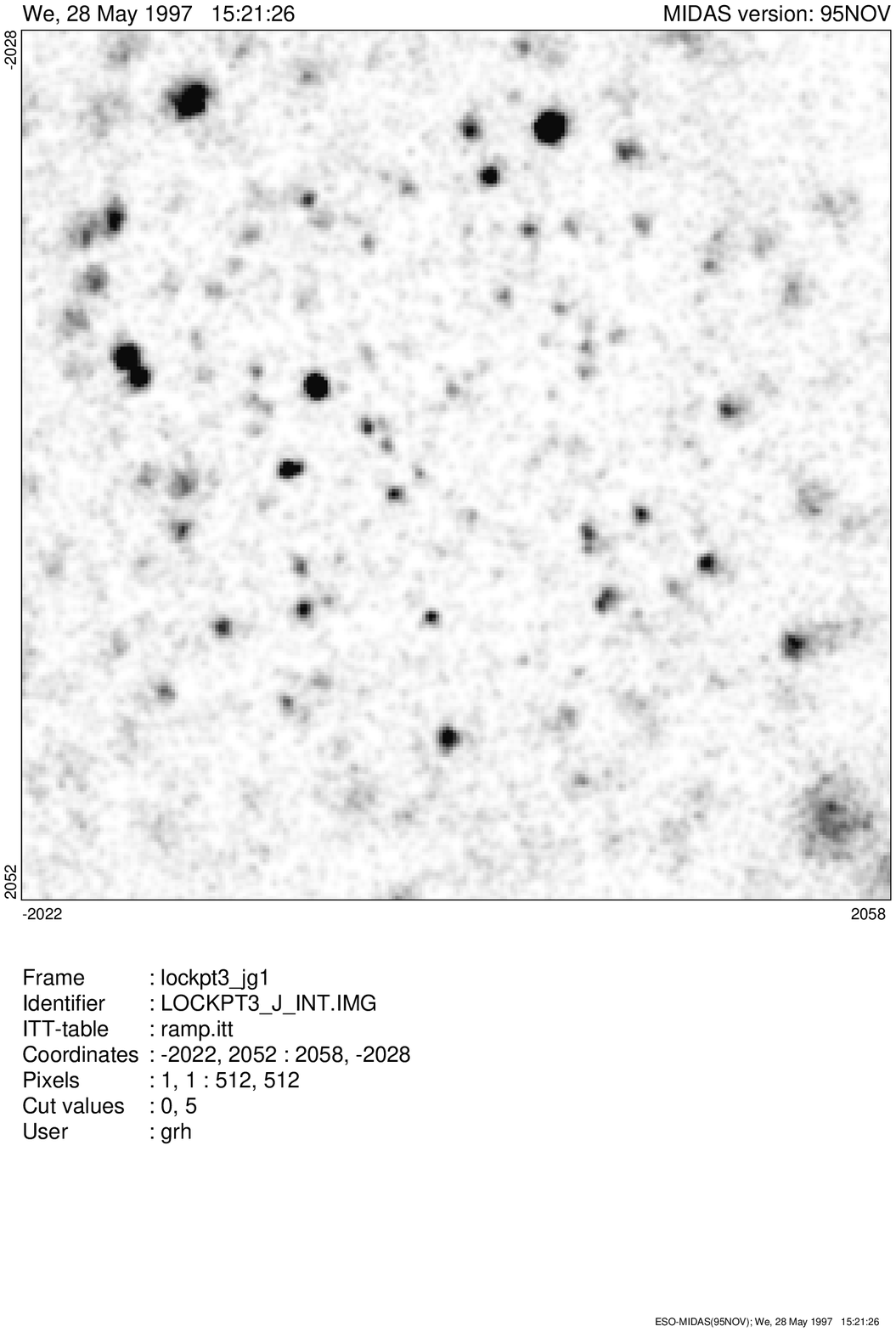,width=5.5cm,clip=}
\end{picture}\par
\end{minipage}
\begin{minipage}[t]{5.5cm}
\begin{picture}(5.5,5.5)
\psfig{figure=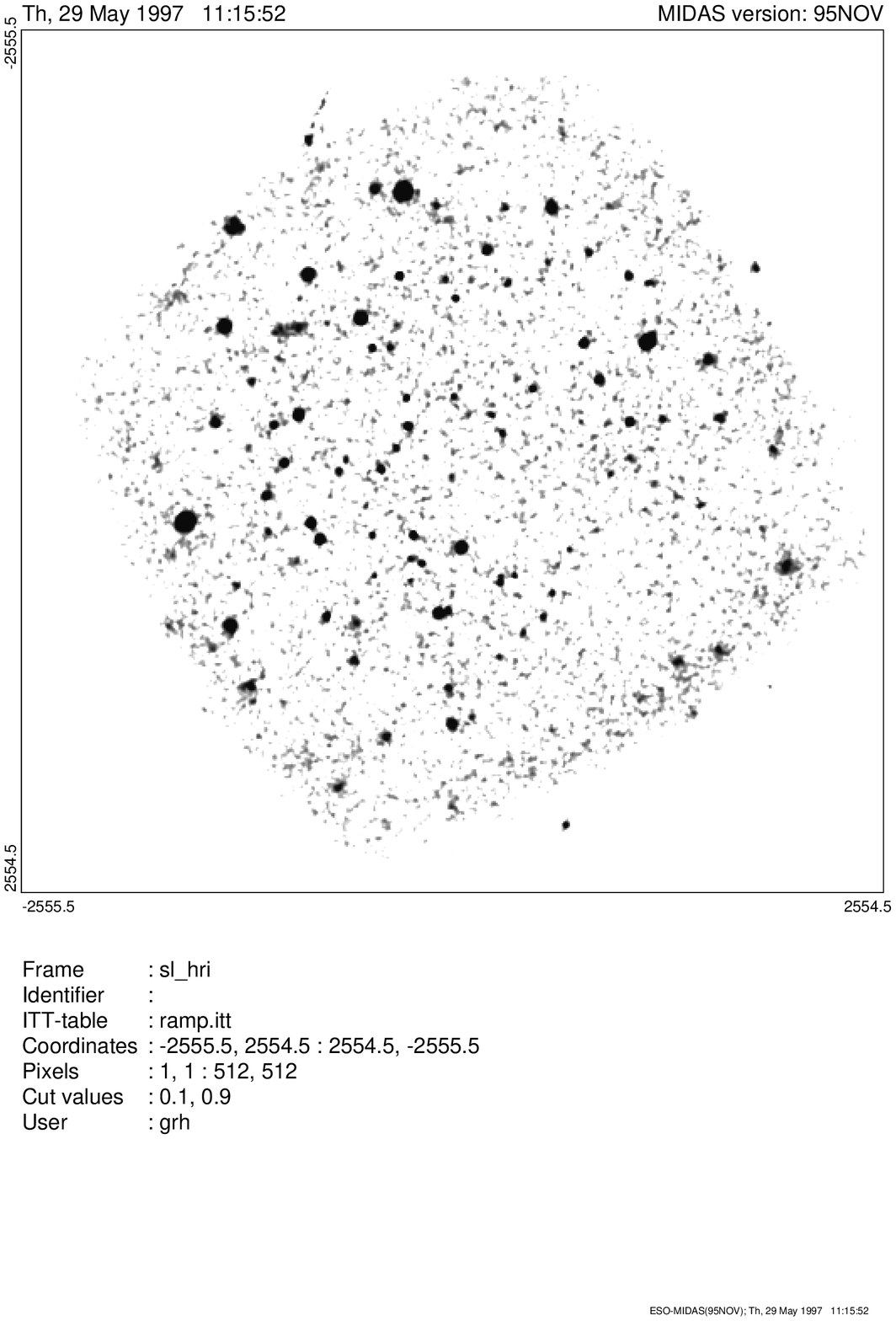,width=5.5cm,clip=}
\end{picture}\par
\vskip -0.2 truecm
\end{minipage}
\caption{\it ROSAT images of the Lockman Hole. Left: 207 ksec PSPC observation,
right: 1 Msec HRI observation. The image size is $34 \times 34$ arcmin.}
\label{IMA}
\end{figure*}

The observations in the {\it Lockman Hole} represent the deepest 
X--ray survey ever performed. The total observing time invested
is quite comparable to that
of other major astronomical projects, like e.g. the Hubble Deep 
Field~\cite{wil96}. 
Figure \ref{IMA} shows the 200 ksec ROSAT PSPC image and the 1 Msec HRI image
of the Lockman Hole. The details of the X--ray observations and data analysis,
as well as a full catalogue of X--ray sources are being published~\cite{has97}.

\subsection{Fighting confusion}

The ROSAT deep survey exposures are probing the limits of observational and
data analysis procedures. In order to obtain a reliable and quantitative
characterization and calibration of the source detection
procedure, detailed simulations of large numbers of artificial fields,
analysed through exactly the same detection and parameter estimation procedure
as the real data, are required. Early simulations~\cite{has93} had already 
demonstrated that source confusion
sets the ultimate limit in ROSAT deep survey work with the PSPC. The
crowded--field multi--ML detection algorithm used in~\cite{has97} was
specifically designed to better cope with source confusion. However, we
felt it necessary to calibrate its efficiency and verify its
accuracy through
new simulations. We have simulated sets of PSPC and HRI observations with
200 ksec and 1 Msec exposure, respectively, approximating our current
observation times. The final results are
relatively independent of the actual log(N)--log(S) parameters
chosen for the artificial fields~\cite{has93}.
In the simulations point sources are placed at random within the field of view,
with fluxes drawn at random from the log(N)--log(S) function
down to a minimum source flux, where
all the X--ray background is resolved for the assumed source counts.
For each source the expected number of counts are computed taking into account
the ROSAT vignetting correction, the appropriate energy to counts conversion
factor and the exposure time of the simulated image.
The actual counts of each source are drawn from a Poissonian distribution
and folded through
the point spread function. The realistic multi--component point spread function
model is taken from~\cite{has94} for the PSPC and from~\cite{dav96} for the HRI.
Finally, all events missing in the field, i.e. particle background and
non--source diffuse background are added as a smooth distribution to the image.
                       
For each detected source the process of ``source identification'' has been
approximated by a simple positional coincidence check. A detected source was
identified with the counterpart from the input list, which appeared closest
to the X--ray position within a radius of 30 arcsec.
The faintest sources detected in the PSPC at small off--axis angles have a flux
of $2 \times 10^{-15}~erg~cm^{-2}~s^{-1}$. 
At larger off--axis angles the sensitivity is
reduced correspondingly. The faintest HRI sources reach down to
fluxes of $10^{-15}~erg~cm^{-2}~s^{-1}$. Again, at larger angles the sensitivity is reduced.

\begin{figure*}[tp]
\unitlength1cm
\begin{minipage}[t]{5.5cm}
\begin{picture}(5.5,5.5)
\psfig{figure=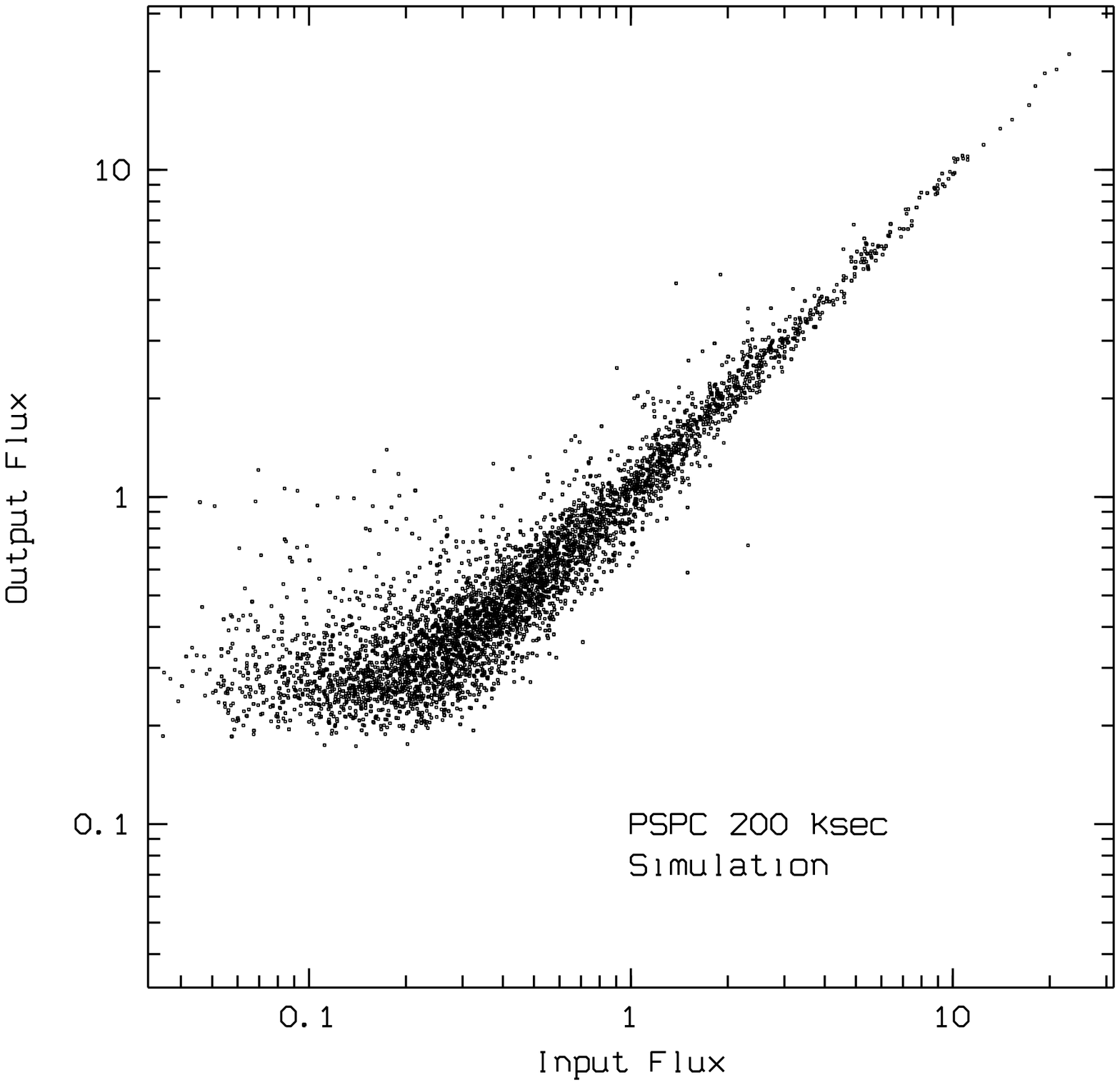,width=6.0cm,angle=-90.}
\end{picture}\par
\end{minipage}
\begin{minipage}[t]{5.5cm}
\begin{picture}(5.5,5.5)
\psfig{figure=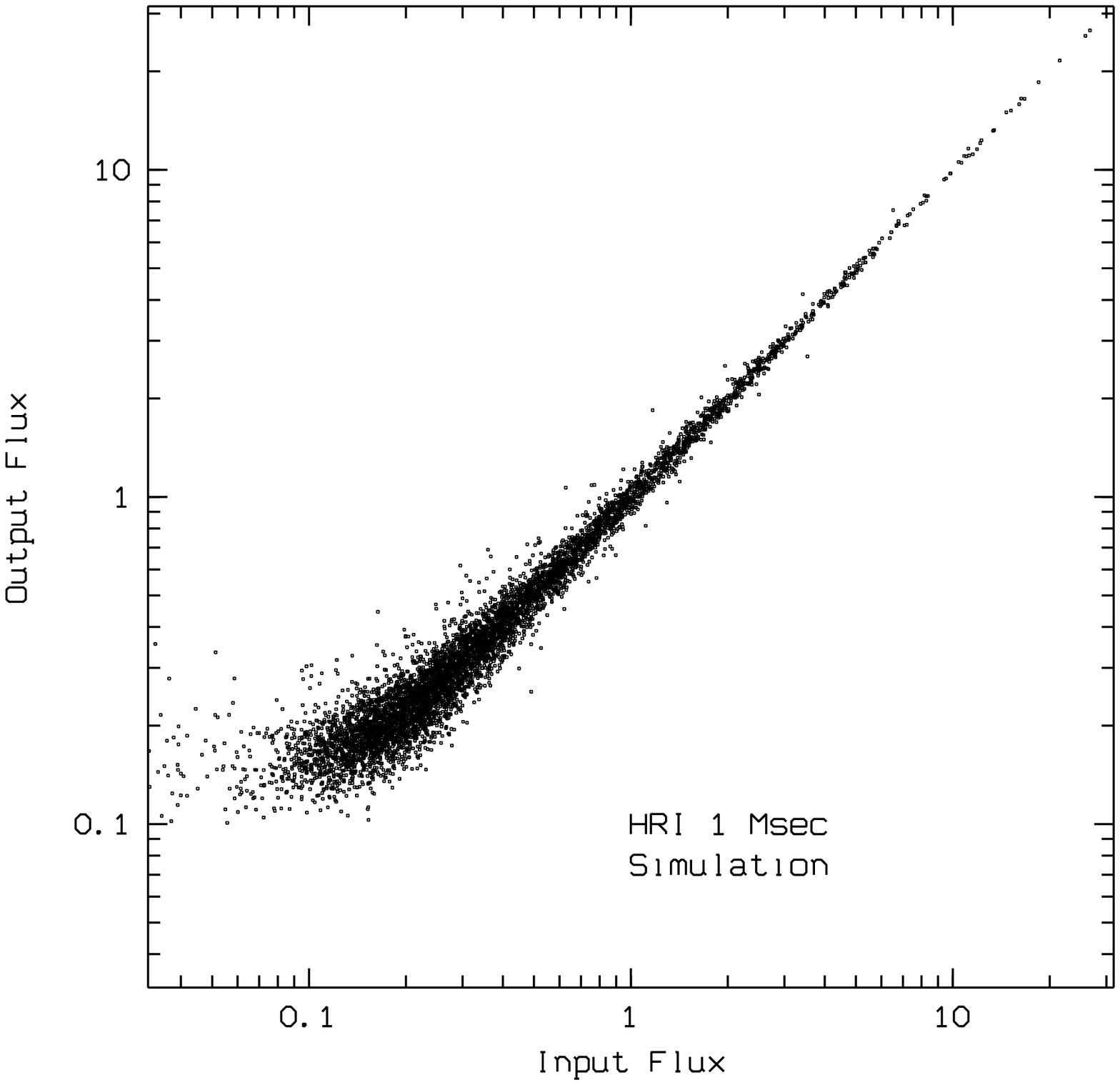,width=6.0cm,angle=-90.}
\end{picture}\par
\vskip -0.2 truecm
\end{minipage}
\caption{\it Detected flux versus input flux for
(a) 66 simulated PSPC fields of 200 ksec exposure, and (b) 100 simulated
HRI fields of 1 Msec exposure each. The PSPC data is for off--axis angles
smaller than 12.5 arcmin, for the HRI the limit is 10 arcmin. Fluxes are in
units of $10^{-14}~erg~cm^{-2}~s^{-1}$.}
\label{FLX}
\end{figure*}

The detected source catalogues are affected by biases and selection effects
present in the source--detection procedure. The most famous of
those is the Eddington bias, which produces a net gain of the number of
sources detected above a given flux limit as a consequence of
statistical errors in the measured flux (see discussion in~\cite{has93}).
Another selection effect, most important in the deep fields considered
here, is source confusion. The net effect of source confusion
is difficult to quantify analytically, because it can affect the
derived source catalogue relation in different ways. Two sub--threshold
sources could be present in the same resolution element and thus
mimic a single detected source. This leads to a net gain
in the number of sources, similar to the Eddington bias.
Two sources above the threshold could merge into a single brighter
source. In this case one source is lost and one is detected at a
higher flux. Whether the total flux is conserved or
not depends on the distance between the two sources and on the details
of the source detection algorithm.
The detection algorithm cannot discriminate close sources with very
different brightness, which results in a net loss of fainter sources.

The effects of confusion become immediately obvious in figure \ref{FLX}
where for each detected source the detected flux is compared to the flux
of the nearest input source within 30 arcsec. While for bright sources
there is an almost perfect match,
there is a significant systematic deviation for fainter X--ray sources, where
most detected objects appear at fluxes significantly brighter than their input
counterparts. This is a direct indication of confusion
because every detected source is only associated with
one input source while its flux may be contributed from several sources.
Source confusion effects are much less pronounced in the HRI data.

While it is obviously possible to correct for confusion effects on the source
counts in a statistical way, the optical identification process relies on the
position of individual sources. Confusion effects like those described here
can cause the position of the detected sources to be significantly offset
with respect to the true source position. This happens more frequently
near the detection limits and therefore complicates the identification
of complete samples of faint sources based on PSPC data alone.

\subsection{The log(N)--log(S) relation}

The first ROSAT deep survey observation, from a $\sim 50~ksec$
PSPC pointing during the verification-phase in the direction of the
North-ecliptic 
pole (NEP), was presented in 1991 in~\cite{has91}.
The log(N)--log(S) function, which reached flux levels about a factor of 
four fainter than the Einstein deep survey limit, clearly showed a 
flattening of the source counts below the Euclidean slope, which was
anticipated previously~\cite{ham92}. The detailed shape
below the break was, however, quite uncertain because of large 
statistical errors. In the following years source counts were presented
from a number of different ROSAT deep surveys, which basically agreed
with the early findings~\cite{has93,bra94,vik95}. 
The best constraints
on the faint-end slope of the source counts is obtained by fluctuation
analyses of PSPC fields with exposure times of 100-150 ksec~\cite{has93,bar94}.

\begin{figure}[tp]
\begin{center}
\unitlength1cm
\begin{minipage}[t]{7.0cm}
\begin{picture}(6.0,7.0)
\centerline{\psfig{figure=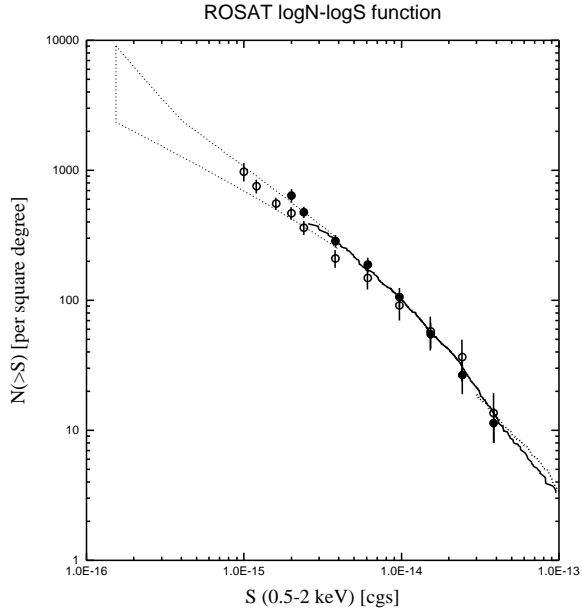,width=7.5cm}}
\end{picture}\par
\end{minipage}
\end{center}
\caption{\it Measured ROSAT log(N)--log(S) function in the Lockman Hole.
Filled circles give the source counts from the 207 ksec PSPC observation.
Open circles are from the ultradeep HRI observation (880 ksec). The data
are plotted on top of the source counts (solid line) and fluctuation
limits (dotted area) from~\protect\cite{has93}. The dotted line at
bright fluxes refers to the total source counts in the RIXOS
survey~\protect\cite{Mas96}.}
\label{LNS}
\end{figure}

The final limiting sensitivity for the detection of discrete sources  
in the Lockman Hole is about $2 \times 10^{-15}~erg~cm^{-2}~s^{-1}$
for the PSPC and about a factor of two fainter with the HRI.
Figure \ref{LNS} shows the new PSPC and HRI log(N)--log(S) 
function~\cite{has97}, 
which is in good agreement with data published previously. The HRI source
counts reaches a surface
density of $970\pm150~deg^{-2}$, about a factor of two higher than any
previous X--ray determination. About 70-80\% of the X--ray background 
measured in the 1-2 keV band has been resolved into discrete sources now.

\subsection{Optical identifications}

Shanks et al.~\cite{sha91} have carried out a program of optical spectroscopy
of sources detected in a 30 ksec PSPC verification phase observation in
one of the AAT deep optical QSO fields and could quickly identify an 
impressive fraction of faint X--ray sources as classical broad-line AGNs (mainly 
QSOs). These studies have 
been continued by the same group on additional fields and deeper 
X--ray data in the following years~\cite{geo96}, until finally 
a large enough sample of AGNs was available to calculate an X--ray 
luminosity function (XLF). Using data from medium-deep ROSAT fields combined
with the Einstein Medium Sensitivity Survey~\cite{gio90}, Boyle et 
al.~\cite{boy94} could improve significantly on the derivation of the AGN XLF and
its cosmological evolution. Their data is consistent with pure 
luminosity evolution proportional to $(1+z)^{2.7}$ up to a redshift 
$z_{max} \approx 1.5$, similar to what was found previously in the optical
range. This result has been confirmed and improved later on by more extensive
or deeper studies of the AGN XLF, e.g. the RIXOS project~\cite{pag96} or the
UK deep survey project~\cite{jon96}. 

All these studies agree that at most half of the faint X--ray source counts 
and, correspondingly, half of the soft X--ray background can be explained
by classical broad-line AGN based on the luminosity evolution models. 
Hasinger et al.~\cite{has93} discussed the possibility that either the XLF models 
have to be modified considerably or a new population of X--ray
sources has to contribute to the X--ray background. 
There was, indeed, mounting evidence that 
a new class of sources might start to contribute to the XRB at faint 
X--ray fluxes. The faintest X--ray sources in ROSAT deep surveys 
on average show a harder spectrum than the identified QSOs~\cite{has93}.
In medium-deep pointings a number of optically ``innocent'' narrow-emission 
line galaxies
(NELGs) at moderate redshifts (z$<$0.4) were identified as X--ray sources, 
which was in excess 
of those expected from spurious identifications with field 
galaxies~\cite{boy95,geo96,gri96}. Roche et al.~\cite{roc95} have 
found a significant correlation of X--ray fluctuations with 
optically faint galaxies. Finally, in an attempt to push optical
identifications to the so far faintest X--ray fluxes, McHardy et 
al.~\cite{mch97} claimed that broad-line AGN practically cease to 
exist at fluxes below $5 \times 10^{-15}~erg~cm^{-2}~s^{-1}$,
while the NELG number counts still keep increasing, so that they would
dominate below a flux of $10^{-15}~erg~cm^{-2}~s^{-1}$.  

While this is obviously an interesting possibility, it is useful to remind that
all these findings are based on identifications near the limit of
deep PSPC surveys, at fluxes where
our simulations suggest that the PSPC data start to be severely confused.
Because moderate-redshift field galaxies in general show 
emission lines, there is the possibility to either misidentify 
a field galaxy as the counterpart of an X--ray source which
in reality is associated to a different optical object (e.g. a fainter AGN)
or to misclassify an intrinsically faint AGN
hidden in a NELG-type spectrum as a new class of X--ray sources (see
also~\cite{has96}). On the contrary,
the X--ray positions in our Lockman Hole survey are largely determined 
by the HRI raster scan and ultradeep pointing. Instead of confused
PSPC error boxes of (realistically) 15-20 arcsec radius, we therefore
have error box radii of 2-5 arcsec. Using optical spectroscopy from the 
Keck telescopes we have recently been able to complete 
the optical identifications in the Lockman Hole down to a flux 
limit of $5.5 \times 10^{-15}~erg~cm^{-2}~s^{-1}$. 
Out of the 50 X--ray sources in the complete X--ray sample,
we identified 38 AGNs, 4 groups of galaxies,
1 normal galaxy and 3 galactic stars. Four X--ray
sources remain unidentified so far, all of which are likely 
to be groups of galaxies~\cite{sch97}. In this survey, which has one of the
highest rates of optical identifications among existing
deep X--ray surveys and has a high degree
of reliability, we see no evidence for the emergence
of a new source population at low X--ray fluxes. We are currently
working on Keck LRIS multi-slit spectroscopy to complete our identifications
down to a flux of $1.5 \times 10^{-15}~erg~cm^{-2}~s^{-1}$ and on the
basis of the data acquired so far we do
not see a need to alter the above conclusions.

\subsection{Constraints from the angular correlation function}

Since, as discussed above, the X--ray background is made up
largely from discrete sources one would expect some variance in the 
spatial distribution of the background due to
these sources. A signal in the angular correlation function (ACF)
can in principle give strong constraints on the
clustering properties of the sources contributing to the X--ray
background.
However, the XRB is remarkably smooth. Until recently
no signal could be detected in the XRB ACF neither at soft X--ray
nor at hard X--ray energies~\cite{fab92}. A significant signal at small
angular separations (1-3 arcmin) could be found 
in a 1-2 keV
analysis of 50 deep ROSAT pointed observations in about $10\%$ of
the fields~\cite{sol94}. However, this signal was 
clearly associated with a few extended, very-low X--ray surface
brightness clusters or groups of galaxies at moderate redshift.
The most 
prominent one of these, the NEP blotch~\cite{has91}, has been in the meantime 
identified with low surface brightness cluster or supercluster 
emission at moderate redshift~\cite{bur92,ash96}. 
In trying to obtain an upper limit on the spatial structure in the background
due to the clustering of sources below the detection limit,
the fields with significant cluster emission have been excluded
from the following analysis. 
Indeed, once those fields were
excluded, only upper limits for the ACF could be obtained~\cite{sol94}. 
However, those limits strongly constrain the nature and clustering
properties of the sources which, being below the detection limit, contribute
to the ``residual'' X--ray
background. According to this analysis, less than $35\%$ of this
residual background can be due to objects with clustering properties
similar to those of bright quasars. The objects which make up the residual
background, i.e. the soft XRB not resolved in the ROSAT
deep surveys, must have a clustering length smaller than normal
galaxies and/or show very strong cosmologic evolution of their
clustering~\cite{sol94}. The smoothness of the XRB therefore provides 
an important additional clue to unravel its composition.

\section{AGN Spectra and Fits to the XRB spectrum} 
The recent significant advances in X--ray spectroscopy made possible
with GINGA, ASCA and BeppoSAX have changed substantially our views on the
spectral characteristics of AGNs and have shown that the X--ray spectra of
AGNs are much more complex than what was thought only a few years ago.
Detailed observations of an increasing number of relatively bright
AGNs have detected a number of spectral features in addition to the
power law continuum, such as the Compton reflection ``hump'' due to reprocessed
emission, absorption from cold material probably in the torus, the warm
absorbers, the Fe lines, the soft excess at low energies (see ~\cite{mu} 
for a recent review). 

Already GINGA data had shown convincingly~\cite{po,na} 
that the typical spectrum of Seyfert 1 galaxies shows a 
flattening at $\sim$ 10 keV, due to the reflected component,
with respect to the observed power law slope in 
the range 2--10 keV. These observations
showed that the average spectrum for these objects is very similar
to the shape of the spectrum hypothesized in~\cite{st}, where
it was shown that such a spectrum,
integrated through redshift with reasonable assumptions on the
cosmological evolution, could provide an adequate fit to the shape
of the observed XRB above 3 keV. 

The Ginga data have immediately led a number of groups to construct models
for fitting the XRB spectrum with various combinations of AGN spectra and
evolution (see, for example,~\cite{mo,fa,te,ro}).
Most of these models require an AGN population whose hard X--ray spectrum
is dominated by the reflected component. Actually, the main difficulty of these
models is the extremely large required contribution of such component.
Moreover, although qualitatively in agreement with the overall
shape of the XRB in the energy range 3--100 keV, it has been shown
in~\cite{zd} that these models are not able to fit
satisfactorily the position 
and the width of the peak of the XRB spectrum.

Alternatively, other studies~\cite{ma,co} have explored in detail
the possibility that
the dominant contribution to the XRB is due to the combination of the
emission from a population of unabsorbed and absorbed AGNs, distributed
according to the unified AGN scheme~\cite{an}, as originally
proposed in~\cite{sw}. This class of models appears to be
highly successful not only in fitting the spectrum of the XRB but also in
reproducing a number of other observational constraints.

\begin{figure}[tp]
\begin{center}
\unitlength1cm
\begin{minipage}[t]{5.5cm}
\begin{picture}(7.5,5.5)
\centerline{\psfig{figure=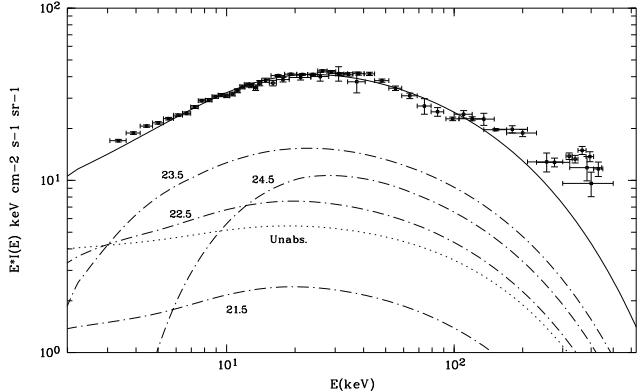,width=8.5cm,angle=-90}}
\end{picture}\par
\end{minipage}
\end{center}
\caption{\it The XRB spectrum above 3 keV: comparison between
model (continuous line) and data. The data above 3 keV are taken from a 
compilation of the best experimental results~\protect\cite{gru}. The curves,
labelled with values of log $N_H$, show the expected contribution from
AGNs with different amount of intrinsic absorption. 
\label{fig:SPECTRUM}
}
\end{figure}

As an example, we show in Figure~\ref{fig:SPECTRUM} the results of a fit to the
XRB spectrum obtained by Comastri {\it et al.}~\cite{co}. The main 
ingredients of this model, which takes into account the observed
spectral properties of different classes of AGNs over a broad energy
range, are the following:

a) The X--ray primary continuum spectrum of AGNs is described by two power laws
with spectral indices $\alpha$ = 1.3 below 1.5 keV and $\alpha$ = 0.9
above this energy. A Compton reflection component (50\% of the primary
spectrum) has been added in the continuum of the low luminosity AGNs~\cite{po} 
(i.e. Seyfert 1 galaxies), while no such
component has been considered in high luminosity AGNs~\cite{wi}. 
In agreement with the OSSE data~\cite{zd95}, the power law 
spectrum is assumed to be exponentially cut--off at an e--folding energy
of about 320 keV. 

b) As required by the adopted unified scheme, the intrinsic X--ray
spectrum of type 2 AGNs is assumed to be the same as the spectrum of type 1
AGNs, but modified by absorption effects~\cite{aw}.
A distribution of $N_H$, i.e. the column density of the
absorbing material, in the range $10^{21} - 10^{25}$ atoms $cm^{-2}$,
consistent with the observed distribution~\cite{scha} is derived.

With these assumptions, the fit shown in Figure~\ref{fig:SPECTRUM} has been
obtained by integrating the local X--ray luminosity function of AGNs
and assuming a luminosity evolution up to z $\sim$ 2. At higher redshift
the X--ray luminosity function has been assumed to remain constant.
This model, as well as other models based on similar assumptions (see, for
example,~\cite{ce}), fits the XRB spectrum well
up to $\sim$ 100 keV, while it underpredicts the data at $\sim$ 500 keV
by a factor of a few. It has been shown in~\cite{co96}
that at this energy the contribution from a different AGN population,
namely flat spectrum radio quasars and ``MeV blazars'', is already substantial.
These objects are likely to contribute most of the observed background in the
energy range $1 - 10^3$ MeV.

As shown in the figure, the model predicts that in the hard X--ray
band most of the contribution to the XRB comes from significantly
absorbed objects, which are almost absent in the soft band, even
at the faintest ROSAT limit. As a consequence, a significant test for this 
model would be the comparison of its predictions with the results
of optical identifications of a complete sample of X--ray sources
selected at low fluxes in the hard X--ray band. 

\begin{figure}[tp]
\begin{center}
\unitlength1cm
\begin{minipage}[t]{6.0cm}
\begin{picture}(6.0,6.0)
\centerline{\psfig{figure=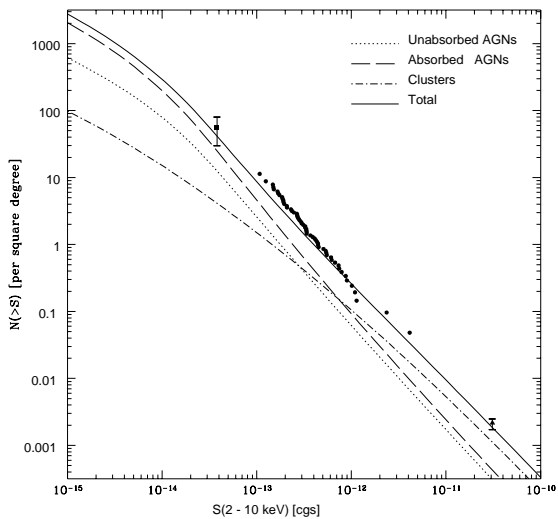,width=7.5cm}}
\end{picture}\par
\end{minipage}
\end{center}
\caption{\it The 2--10 keV log(N)--log(S). The circles represent the counts
derived from the 60 ASCA sources discussed in~\protect\cite{cdm}; the square
shows the preliminary results of a different ASCA survey in a more limited sky
area~\protect\cite{oga}; the triangle shows the extragalactic surface
density from the HEAO-1 survey~\protect\cite{pic}. The curves show 
the predictions for various classes of objects from
the model discussed in ~\protect\cite{co}. 
\label{fig:ASCANS}
}
\end{figure}

Recently, ASCA
data have been used~\cite{cdm} to derive  
the 2--10 keV log(N)--log(S) down to fluxes slightly below $10^{-13}
erg~cm^{-2}~s^{-1}$. This data, based on 60 X--ray sources detected
in ASCA images, is shown in Figure~\ref{fig:ASCANS} together with the
predictions for various classes of objects from
the Comastri {\it et al.} model. As discussed in Cagnoni 
{\it et al.}~\cite{cdm},  
if one uses the ROSAT log(N)--log(S) and the average spectral properties
of the ROSAT sources to predict the ASCA log(N)--log(S), the prediction
falls a factor $\sim$ 2 below the data at a flux of the order
of $\sim$ $10^{-13} erg~cm^{-2}~s^{-1}$. On the other hand, as shown in
the figure, the data are in good agreement with the predictions 
of the Comastri {\it et al.} model. The dashed line in Figure~\ref{fig:ASCANS}
shows that a significant fraction
of the AGNs in this ASCA survey, and an even larger fraction at fainter
fluxes, should be substantially
absorbed and therefore their optical counterparts are expected to have
optical spectra typical of Seyfert 2 galaxies. A program to optically identify
these sources has already started, but the large positional uncertainty 
($\sim$ 2 arcmin) together with the relatively faint expected
magnitudes of the optical counterparts will make it difficult to obtain
unambiguous identifications. Fortunately, the situation will soon improve
with the missions to be launched in the next few years, as for example 
AXAF and XMM. These missions will provide much fainter samples of hard
X--ray selected sources, with significantly better positional uncertainty
(of the order of 1--2 arcsec for AXAF and 5--10 arcsec for XMM). These faint 
hard samples will be complemented by the bright all-sky sample from the 
ABRIXAS survey.
The optical identifications of these sources with the 8m class telescopes
will be able to conclusively test the predictions of our current models
and, hopefully, finally settle the question of the X-ray background.

\section{Conclusions}

In the last decade there has been a substantial
improvement in our understanding of both the 
X-ray background and the spectra of active galactic nuclei. To first
order we can regard the problem of the X-ray background as solved. For
the first time since its discovery we can explain the bulk of the X-ray 
background over a wide range of energies (0.1-100 keV) as due to a known class
of sources, namely unabsorbed and absorbed active galactic nuclei, whose 
statistical properties
like average spectra, absorption distribution, luminosity function etc. are 
consistent with those
measured locally and which are strongly evolving with cosmic time.

In the soft X-ray band, mainly based on ROSAT deep surveys, the majority 
of identified sources down to the faintest fluxes are indeed AGNs and the
background models can account for all observable constraints
like log(N)-log(S) function, redshift distribution, average spectral index
etc.
Observations at harder X-ray energies, e.g. by ASCA, are consistent with
this picture. However, only deeper survey observations with the next generation
of sensitive X-ray observatories with good angular resolution in the 
harder X-ray band (i.e. AXAF and XMM) and optical follow-up spectroscopy from
8-10m telescopes will be able to unambiguously confirm the AGN XRB model and to 
tighten the constraints on some of the still relatively uncertain 
model parameters.

\section*{Acknowledgements}
{\small G.H. acknowledges the grant 50 OR 9403 5 by the Deutsche Agentur f\"ur
Raumfahrtangelegenheiten (DARA) and G.Z. the Italian Spqce Agency (ASI) contract
ASI 95-RS-152. We also warmly thank all our collaborators in the ROSAT Deep Survey
project and A. Comastri for fruitful discussions.


\section*{References}
\small
\begin{thebibliography}{99}
\bibitem{gb}    Giacconi R. and Burg R.  1992, in {\it The X--Ray 
                Background}, X. Barcons and A.C. Fabian eds., (Cambridge: 
                Cambridge Univ. Press), 3.
\bibitem{rg62}  Giacconi R., Gursky H., Paolini F.R. and Rossi B.B. 1962, 
                {\it Phys.Rev.Letters}, {\bf 9}, 439.
\bibitem{sc}    Schwartz D.A. 1980, {\it Phys.Scripta}, {\bf 21}, 644.
\bibitem{ta}    Tanaka Y. 1992, in {\it X--Ray Emission from Active Galactic
                Nuclei and the Cosmic X--Ray Background}, W. Brinkmann and J. 
                Tr\"umper eds., MPE Report {\bf 235}, 303.
\bibitem{mar}   Marshall F.E. {\it et al.} 1980, {\it Ap.J.}, {\bf 235}, 4.
\bibitem{mu84}  Mushotzky R.F. 1984, {\it Advances in Space Research}, Vol.
                {\bf 3}, no. 10--12, p.157.
\bibitem{sw73}  Setti G. and Woltjer, L. 1973, in IAU Symposium No. 55,
                {\it X- and Gamma--Ray Astronomy}, H. Bradt and R. Giacconi 
                eds., (Reidel, Dordrecht), p. 187
\bibitem{ht}    Tananbaum R. {\it et al.} 1979, {\it Ap.J.Letters}, {\bf 234},
                L9.
\bibitem{gz81}  Zamorani G. {\it et al.} 1981, {\it Ap.J.}, {\bf 245}, 357.
\bibitem{rg79}  Giacconi R. {\it et al.} 1979, {\it Ap.J.Letters}, {\bf 234}, 
                L1.
\bibitem{gr}    Griffiths R.E. {\it et al.} 1983, {\it Ap.J.}, {\bf 269}, 375.
\bibitem{pr}    Primini F.A. {\it et al.} 1991, {\it Ap.J.}, {\bf 374}, 440.
\bibitem{ha}    Hamilton T.T., Helfand D.J. and Wu X. 1991, {\it Ap.J.}, {\bf
                379}, 576.
\bibitem{se}    Setti G. 1985, in {\it Non--Thermal and Very High Temperature
                Phenomena in X--Ray Astronomy}, G.C. Perola and M. Salvati 
                eds., p. 159.
\bibitem{gz}    Giacconi R. and Zamorani G. 1987, {\it Ap.J.}, {\bf 313}, 20.
\bibitem{mat90} Mather J.C. {\it et al.} 1990, {\it Ap.J.Letters}, {\bf 354},
                L37.
\bibitem{mat94} Mather J.C. {\it et al.} 1994, {\it Ap.J.}, {\bf 420}, 439.
\bibitem{wr}    Wright {\it et al.} 1994, {\it Ap.J.}, {\bf 420}, 450.
\bibitem{pfe86} Pfeffermann E. {\it et al.} 1986, {\it SPIE}, {\bf 733}, 519.
\bibitem{sol91} Soltan A.M. 1991, {\it MNRAS}, {\bf 250}, 241.
\bibitem{ham92} Hamilton T.T. 1992, in {\it The X--Ray Background}, 
                X. Barcons and A.C. Fabian eds., (Cambridge: Cambridge Univ. 
                Press), 138.
\bibitem{bri88} Briel U.G. {\it et al.} 1988, {\it SPIE}, {\bf 982}, 401.
\bibitem{har86} Hartner G. 1986, Diplomarbeit TU M\"unchen
\bibitem{has93} Hasinger G. {\it et al.} 1993, {\it A\&A} 
                {\bf 275}, 1.
\bibitem{loc86} Lockman F.J., Jahoda K. and McCammon D. 1986, {\it Ap.J.},
                {\bf 302}, 432.
\bibitem{mar88} Marano B., Zamorani G. and Zitelli V. 1988, {\it MNRAS},
                {\bf 212}, 111.
\bibitem{sgr83} Schmidt M. and Green R.P. 1983, {\it Ap.J.}, {\bf 269}, 352.
\bibitem{der97} de Ruiter H. {\it et al.} 1997, {\it A\&A}, {\bf 319}, 7.
\bibitem{has94} Hasinger G. {\it et al.} 1994, {\it Legacy}, {\bf 4}, 40; 
                MPE/OGIP Calibration Memo.
\bibitem{dav96} David F.R. {\it et al.} 1996, in {\it The ROSAT Users Handbook}
\bibitem{wil96} Williams R.E. {\it et al.} 1996. {\it A.J.}, {\bf 112}, 
                1335.
\bibitem{has97} Hasinger {\it et al.} 1997, {\it A\&A}, submitted.
\bibitem{has91} Hasinger G., Schmidt M. and Tr\"umper J. 1991, {\it A\&A}, 
                {\bf 246}, L2.
\bibitem{bra94} Branduardi-Raymont G. {\it et al.}  1994, {\it MNRAS}, 
                {\bf 270}, 947.
\bibitem{vik95} Vikhlinin A. {\it et al.} 1995, {\it Ap.J.}, {\bf 451}, 553.
\bibitem{bar94} Barcons X. {\it et al.} 1994, {\it MNRAS}, {\bf 268}, 833.
\bibitem{Mas96} Mason, K. {\it et al.} 1996, priv. comm.
\bibitem{sha91} Shanks T. {\it et al.} 1991, {\it Nature}, {\bf 353}, 315.
\bibitem{geo96} Georgantopoulos I. {\it et al.} 1996, {\it MNRAS}, 
                {\bf 280}, 276.
\bibitem{gio90} Gioia I.M. {\it et al.} 1990, {\it Ap.J.Suppl.}, {\bf 72}, 567.
\bibitem{boy94} Boyle B.J. {\it et al.} 1994, {\it MNRAS}, {\bf 260}, 49.
\bibitem{pag96} Page M.J. {\it et al.} 1996, {\it MNRAS}, {\bf 281}, 579.
\bibitem{jon96} Jones L.R. {\it et al.} 1996, {\it MNRAS}, submitted,
                (astro-ph/9610124).
\bibitem{boy95} Boyle B.J. {\it et al.} 1995, {\it MNRAS}, {\bf 272}, 462.
\bibitem{gri96} Griffiths R.E. {\it et al.} 1996, {\it MNRAS}, {\bf 281}, 
                71.
\bibitem{roc95} Roche N. {\it et al.} 1995, {\it MNRAS}, {\bf 273}, L15.
\bibitem{mch97} McHardy I.M. {\it et al.} 1997, {\it MNRAS}, submitted, 
                (astro-ph/9703163)
\bibitem{has96} Hasinger G. 1996, {\it A\&A Suppl.}, {\bf 120}, C607.
\bibitem{sch97} Schmidt M. {\it et al.} 1997, {\it A\&A}, submitted.
\bibitem{fab92} Fabian A.C. and Barcons X. 1992, {\it ARA\&A}, {\bf 30}, 429.
\bibitem{sol94} Soltan A. and Hasinger G. 1994, {\it A\&A}, 288, 77.
\bibitem{bur92} Burg R. {\it et al.} 1992, {\it A\&A}, {\bf 259}, L9.
\bibitem{ash96} Ashby M.N.L. {\it et al.} 1996, {\it Ap.J.}, {\bf 456}, 428.
\bibitem{mu}    Mushotzky R.F. 1997, in {\it Mass Ejection from AGN},
                in press, (astro-ph/9705004). 
\bibitem{po}    Pounds K.A. {\it et al.} 1990, {\it Nature}, {\bf 344}, 132.
\bibitem{na}    Nandra K. 1991, {\it Ph.D. Thesis}, Leicester University.
\bibitem{st}    Schwartz D.A. and Tucker W.H. 1988, {\it Ap.J.}, {\bf 332},
                157.
\bibitem{mo}    Morisawa K., Matsuoka M., Takahara F. and Piro L. 1990,
                {\it A\&A}, {\bf 263}, 299.
\bibitem{fa}    Fabian A.C., George I.M., Miyoshi S. and Rees M.J. 1990, 
                {\it MNRAS}, {\bf 242}, 14P.
\bibitem{te}    Terasawa N. 1991, {\it Ap.J.Letters}, {\bf 378}, L11.
\bibitem{ro}    Rogers R.D. and Field G.B. 1991, {\it Ap.J.Letters}, 
                {\bf 370}, L57.
\bibitem{zd}    Zdziarski A.A., Zycki P.T., Svensson R. and Boldt E. 1993,
                {\it Ap.J.}, {\bf 405}, 125.
\bibitem{ma}    Madau P., Ghisellini G. and Fabian A.C. 1994, {\it MNRAS},
                {\bf 270}, 117.
\bibitem{co}    Comastri A., Setti G., Zamorani G. and Hasinger G. 1995, 
                {\it A\&A}, {\bf 296}, 1.
\bibitem{an}    Antonucci R.J. 1995, {\it ARA\&A}, {\bf 31}, 473.
\bibitem{sw}    Setti G. and Woltjer L. 1989, {\it A\&A}, {\bf 224}, L21.
\bibitem{gru}   Gruber D.E. 1992,  in {\it The X--Ray Background}, 
                X. Barcons and A.C. Fabian eds., (Cambridge: Cambridge Univ. 
                Press), 44.
\bibitem{wi}    Williams O.R. {\it et al.} 1992, {\it Ap.J.}, {\bf 389}, 157.
\bibitem{zd95}  Zdziarski A.A. {\it et al.} 1995, {\it Ap.J.Letters}, 
                {\bf 438}, L63.
\bibitem{aw}    Awaki H., Koyama K., Inoue H. and Halpern J.P. 1991, {\it 
                PASJ}, {\bf 43}, 195.
\bibitem{scha}  Schartel N. {\it et al.} 1997, {\it A\&A} {\bf 320}, 696.
\bibitem{ce}    Celotti A., Fabian A.C., Ghisellini G. and Madau P. 1996,
                {\it MNRAS}, {\bf 277}, 1169.
\bibitem{co96}  Comastri A., Di Girolamo T. and Setti G. 1996, 
                {\it A\&A Suppl.}, {\bf 120}, C627.
\bibitem{cdm}   Cagnoni I., Della Ceca R. and Maccacaro T. 1997, 
                {\it Ap.J.}, submitted.
\bibitem{oga}   Ogasaka Y. 1997, in {\it X--ray Imaging and Spectroscopy
                of Cosmic Hot Plasmas}, F. Makino and K. Mitsuda eds., 25.
\bibitem{pic}   Piccinotti G. {\it et al.} 1982, {\it Ap.J.}, {\bf 253}, 485.
\end{thebibliography}
\end{document}